\documentclass[11pt]{article}%
\usepackage{amssymb}
\usepackage{amsfonts}
\usepackage{amsmath}
\usepackage{graphicx}
\usepackage{algorithmic}
\usepackage{float}
\usepackage[ruled,nothing]{algorithm}
\usepackage{subfig}%
\providecommand{\U}[1]{\protect\rule{.1in}{.1in}}
\newtheorem{theorem}{Theorem}

\newtheorem{definition}[theorem]{Definition}

\newtheorem{lemma}[theorem]{Lemma}

\newdimen\dummy
\dummy=\oddsidemargin
\begin{document}

\title{Modeling Social Networks with Overlapping Communities Using Hypergraphs and
Their Line Graphs}
\author{D. Liu, N. Blenn and P. Van Mieghem\thanks{Faculty of Electrical Engineering,
Mathematics and Computer Science, Delft University of Technology, P.O Box
5031, 2600 GA Delft, The Netherlands; \emph{email}:
\{D.Liu,N.Blenn,P.F.A.VanMieghem\}@tudelft.nl. }}
\date{Delft University of Technology}
\maketitle

\begin{abstract}
We propose that hypergraphs can be used to model social networks with
overlapping communities. The nodes of the hypergraphs represent the
communities. The hyperlinks of the hypergraphs denote the individuals who may
participate in multiple communities. The hypergraphs are not easy to analyze,
however, the line graphs of hypergraphs are simple graphs or weighted graphs,
so that the network theory can be applied. We define the overlapping depth $k$
of an individual by the number of communities that overlap in that individual,
and we prove that the minimum adjacency eigenvalue of the corresponding line
graph is not smaller than $-k_{\max}$, which is the maximum overlapping depth
of the whole network. Based on hypergraphs with preferential attachment, we
establish a network model which incorporates overlapping communities with
tunable overlapping parameters $k$ and $w$. By comparing with the Hyves social
network, we show that our social network model possesses high clustering,
assortative mixing, power-law degree distribution and short average path length.

\end{abstract}

\floatstyle{ruled} \newfloat{algorithm}{htbp}{loa}
\floatname{algorithm}{Algorithm}
%\floatname{algorithm}{Procedure}

%\pagestyle{empty}

\section{Introduction}

Social networks, as one type of real-world complex networks, are currently
widely studied
\cite{Albert:statistical_mechanics_networks,Boccaletti:complex_networks_structure_dynamics,Newman_randomgraph,Girvan:socialnet_communities}%
. Most social networks have common properties of the real-world networks, such
as high clustering coefficient, short characteristic path length, power law
degree distribution
\cite{Albert:statistical_mechanics_networks,Newman_randomgraph,BApowerlaw,Watts-Strogatz}%
. Meanwhile, they possess some special properties like assortative mixture,
community and hierarchical structure
\cite{Girvan:socialnet_communities,Ahn:nature_link_communities,newman:mixing_patterns,Piet:influence_assortativity}%
. The communities are the subunits of a network, which exhibit relatively
higher levels of connections within the subunits and a lower connectivity
between the subunits. Community structures feature important topological
properties that have catalyzed researches on communities detection algorithms
and on modularity analysis
\cite{Fortunato201075:communities_detection_review,newman:finding_communities,Piet:Spec_Modularity}%
. The communities overlap with each other when nodes belong to multiple
communities. The overlap of different communities exists naturally in
real-world complex networks, particularly in social and biological networks
\cite{Palla:overlapping_communities,Evans_2,McDaid:detecting_overlapping_commutities}%
. The overlap is present at the interface between communities and could also
be pervasive in the whole network. The existence of overlapping communities
challenge the traditional algorithms and methods
\cite{Fortunato201075:communities_detection_review} for community detection
and network (nodes) partitioning. Ahn et al.
\cite{Ahn:nature_link_communities} and Evans et al.
\cite{PhysRevE.80.016105_linkPartition} proposed that partitioning the links
of the concerned network could be void of overlapping communities. Actually
this method only works when two communities overlap at most in one node, as
shown in Figure \ref{Fig_overlapping_depth_width} (a). If two communities
overlap in two or more nodes, they also overlap in links, as shown in Figure
\ref{Fig_overlapping_depth_width} (b)\ where the thick black links belong to
two communities.

We propose that hypergraphs and line graphs of hypergraphs can be used to
model the networks with overlapping communities. A hypergraph is the
generalization of a simple graph. A hypergraph $H\left(  N,L\right)  $ has the
same types of nodes as a simple graph \cite{Harary}, but its
hyperlinks\footnote{The hyperlinks here should not be confused with hyperlinks
of WWW webs. Some papers call them hyperedges.} can connect a variable number
$k$ of nodes, $k=1,2,3,\cdots$. Here $N$ and $L$ denote the number of nodes
and hyperlinks respectively. The line graph $l\left(  H\right)  $ of a
hypergraph $H\left(  N,L\right)  $\ is a graph in which every node of
$l\left(  H\right)  $ represents a hyperlink of $H\left(  N,L\right)  $ and
two nodes of $l\left(  H\right)  $ are adjacent if and only if their
corresponding links share node(s) in $H\left(  N,L\right)  $
\cite{Line_graphs_of_hypergraphs_I}. As discussed in Section
\ref{section_lineG_hypergraph}, the line graph $l\left(  H\right)  $ is a
simple graph\footnote{A simple graph is an unweighted, undirected graph
containing no self-loops (links starting and ending at the same node) nor
multiple links between the same pair of nodes.} when $H\left(  N,L\right)  $
is linear\footnote{A hypergraph is linear if each pair of hyperlinks share at
most one node. Hypergraphs where all hyperlinks connect the same number $k$ of
nodes are defined as $k$-uniform hypergraphs. A $2$-uniform hypergraph is a
simple graph.}, otherwise $l\left(  H\right)  $ is a weighted graph. Applying
the concepts to communities, we have that:

\begin{itemize}
\item Hypergraphs: The nodes represent the communities; The hyperlinks denote
the individuals who may belong to multiple communities. If an individual
belongs to several communities, the corresponding nodes are connected by the
corresponding hyperlink.

\item Line graphs of hypergraphs: The nodes represent the individuals. The
communities consist of the participating nodes and all the links
inter-connecting them. Two individuals are connected by a link if they belong
to the same community. All the communities are the cliques in the line graph.
\end{itemize}

By using hypergraphs and their line graphs, we establish in this article a
network model which incorporates overlapping communities structures for the
first time with tunable overlapping parameters: the overlapping depth $k$ and
the overlapping width $w$ (defined in Section
\ref{section_overlapping_parameters}). By introducing the preferential
attachment to hypergraphs, we obtain a power-law community size distribution
and a power-law degree distribution. Our network model also possesses high
clustering, assortative mixing and short average path length. We compare the
mentioned metrics of our model with the corresponding metrics of an online
social network retrieved from a part of public profiles of Hyves, which is the
popular Dutch social networking site.

\section{Hypergraphs modeling social networks with overlapping communities}

\subsection{The overlapping parameters for communities}

\label{section_overlapping_parameters}Human beings have multiple roles in the
society, and these roles make people members of multiple communities at the
same time, such as companies, universities, families/relationships, hobby
clubs, etc. Proteins may also involve in multiple functional categories in
biological networks. That is how overlapping communities emerge in social and
biological networks. Sometimes only two communities overlap in the same node,
and sometimes a huge number of communities overlap in the same node. Two
communities may overlap only in one node and they may also overlap in many nodes.

\begin{definition}
We define the overlapping depth $k$ of an individual by the number of
communities that overlap in that individual.
\end{definition}

\begin{definition}
We define the overlapping width $w$ of two communities by the number of
individuals that they overlap.
\end{definition}

\begin{figure}
[ptb]
\begin{center}
\includegraphics[
height=2.2459in,
width=4.7392in
]%
{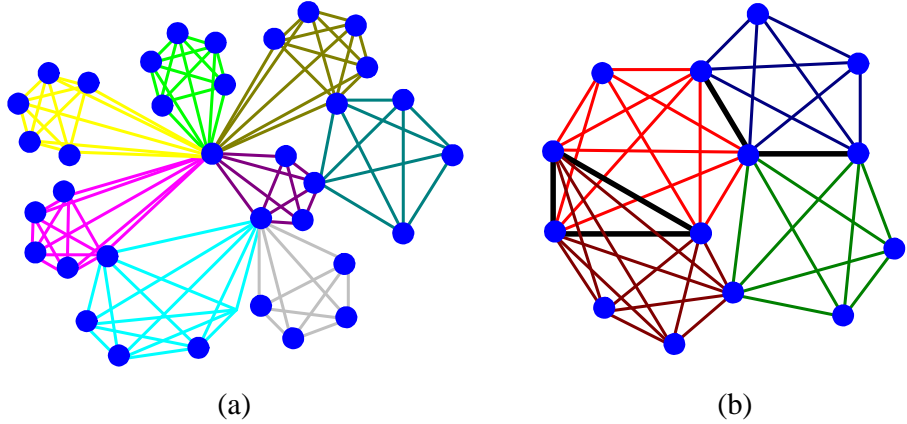}%
\caption{The example graphs showing the overlapping depths of nodes and the
overlapping widths of two communities. The nodes denote the individuals. The
communities consist of links of the same color and the shared thick black
link(s) and the nodes incident to the links.}%
\label{Fig_overlapping_depth_width}%
\end{center}
\end{figure}

The nodes in Figure \ref{Fig_overlapping_depth_width}\ denote the individuals.
There are five individuals in Figure \ref{Fig_overlapping_depth_width} (a)
which have at least two communities overlapping in them. The overlapping
depths of them are $5,3,2,2,2$. As shown in Figure
\ref{Fig_overlapping_depth_width} (b), the overlapping widths of four
community pairs, red and brown, red and dark blue, green and dark blue, brown
and green, are $3,2,2,1$. The individuals of the social network modeled by a
$k$-uniform hypergraph all belong to $k$ different communities, hence, the
overlapping depths of all hyperlinks of a $k$-uniform hypergraph are $k$. The
overlapping width of any node pair of a linear hypergraph is not larger than
$1$, regarding nodes as communities and hyperlinks as individuals,
%TCIMACRO{\FRAME{ftbpFU}{4.7392in}{2.2459in}{0pt}{\Qcb{The example graphs
%showing the overlapping depths of nodes and the overlapping widths of two
%communities. The nodes denote the individuals. The communities consist of
%links of the same color and the shared thick black link(s) and the nodes
%incident to the links.}}{\Qlb{Fig_overlapping_depth_width}}%
%{overlapping_depth_width.eps}{\special{ language "Scientific Word";
%type "GRAPHIC";  maintain-aspect-ratio TRUE;  display "USEDEF";
%valid_file "F";  width 4.7392in;  height 2.2459in;  depth 0pt;
%original-width 4.7124in;  original-height 2.2182in;  cropleft "0";
%filename 'overlapping_depth_width.eps';file-properties "XNPEU";}}}%
%BeginExpansion
%EndExpansion

\subsection{Modeling social networks}

\begin{figure}
[ptb]
\begin{center}
\includegraphics[
height=2.4872in,
width=4.3232in
]%
{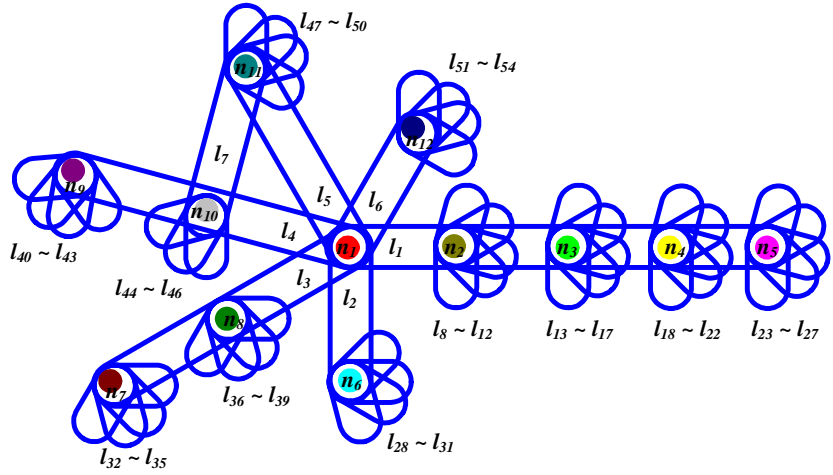}%
\caption{An example of a hypergraph modeling a small size real social network.
The hyperlinks and nodes represent the individuals and the communities
respectively. Each individual may participate in multiple communities, in
other words, the communites overlap with each other.}%
\label{hypergraph_socialnetworks}%
\end{center}
\end{figure}

The hyperlinks and nodes represent the individuals and the communities
respectively. People may participate in multiple communities. If an individual
belongs to several communities, the corresponding nodes are connected by the
corresponding hyperlink. We show how a hypergraph models a real social network
by an example of Figure \ref{hypergraph_socialnetworks}. This is a small
social network of a research group NAS\footnote{Network Architectures and
Services group} at TU Delft. Despite of its small size, the overlapping
communities still emerge. In Figure \ref{hypergraph_socialnetworks}, there are
$12$ communities as described in Table \ref{Tbl_nodes_communites}, and there
are $54$ individuals among whom $6$ individuals belong to NAS group possessing
overlapping depth of $5,3,3,2,2,2$. The 7th individual joins in both the
communities of a rock band and a soccer team.

The hypergraphs are too complicated to implement network analysis, however,
the line graphs of hypergraphs are simple graphs or weighted graphs whose
properties are easier to investigate.%

%TCIMACRO{\FRAME{ftbpFU}{4.3232in}{2.4872in}{0pt}{\Qcb{An example of a
%hypergraph modeling a small size real social network. The hyperlinks and nodes
%represent the individuals and the communities respectively. Each individual
%may participate in multiple communities, in other words, the communites
%overlap with each other.}}{\Qlb{hypergraph_socialnetworks}}%
%{hypergraph_socialnetworks_1.eps}{\special{ language "Scientific Word";
%type "GRAPHIC";  maintain-aspect-ratio TRUE;  display "USEDEF";
%valid_file "F";  width 4.3232in;  height 2.4872in;  depth 0pt;
%original-width 3.9695in;  original-height 2.4491in;  cropleft "0";
%filename 'hypergraph_socialnetworks_1.eps';file-properties "XNPEU";}}}%
%BeginExpansion
%EndExpansion%
%TCIMACRO{\TeXButton{B}{\begin{table}[tbp] \centering}}%
%BeginExpansion
\begin{table}[tbp] \centering
%EndExpansion%
\begin{tabular}
[c]{|l|l|}\hline
\textbf{Nodes} & \textbf{Communities}\\\hline
$n_{1}$ & TU Delft research group-NAS\\\hline
$n_{2}$ & MIT research group\\\hline
$n_{3}$ & Cornell Univ. research group\\\hline
$n_{4}$ & IEEE/ACM ToN editorial board\\\hline
$n_{5}$ & Kansas State Univ. research group\\\hline
$n_{6}$ & Ericsson\\\hline
$n_{7}$ & KPN (Dutch Telecom)\\\hline
$n_{8}$ & Piano club\\\hline
$n_{9}$ & TNO (A Dutch consulting company)\\\hline
$n_{10}$ & A rock band\\\hline
$n_{11}$ & A soccer team\\\hline
$n_{12}$ & TU Delft research group-Bioinformatics\\\hline
\end{tabular}
%TCIMACRO{\TeXButton{caption}{\caption
%{The details of all communities of the NAS social network.}}}%
%BeginExpansion
\caption{The details of all communities of the NAS social network.}%
%EndExpansion
\label{Tbl_nodes_communites}%
%TCIMACRO{\TeXButton{E}{\end{table}}}%
%BeginExpansion
\end{table}%
%EndExpansion%
%TCIMACRO{\TeXButton{B}{\begin{table}[tbp] \centering}}%
%BeginExpansion
\begin{table}[tbp] \centering
%EndExpansion%
\begin{tabular}
[c]{|l|l|}\hline
\textbf{Communities} & \textbf{Individuals}\\\hline
$n_{1}$ & $l_{1}$ to $l_{6}$\\\hline
$n_{2}$ & $l_{1},l_{8}$ to $l_{12}$\\\hline
$n_{3}$ & $l_{1},l_{13}$ to $l_{17}$\\\hline
$n_{4}$ & $l_{1},l_{18}$ to $l_{22}$\\\hline
$n_{5}$ & $l_{1},l_{23}$ to $l_{27}$\\\hline
$n_{6}$ & $l_{2},l_{28}$ to $l_{31}$\\\hline
$n_{7}$ & $l_{3},l_{32}$ to $l_{35}$\\\hline
$n_{8}$ & $l_{3},l_{36}$ to $l_{39}$\\\hline
$n_{9}$ & $l_{4},l_{40}$ to $l_{43}$\\\hline
$n_{10}$ & $l_{4},l_{7},l_{44}$ to $l_{46}$\\\hline
$n_{11}$ & $l_{5},l_{7},l_{47}$ to $l_{50}$\\\hline
$n_{12}$ & $l_{6},l_{51}$ to $l_{54}$\\\hline
\end{tabular}
\caption{The members of all the communities of the NAS social network.}\label{Tbl_communities_nodes_linegraph}%
%TCIMACRO{\TeXButton{E}{\end{table}}}%
%BeginExpansion
\end{table}%
%EndExpansion

\section{The line graphs of hypergraphs}

\label{section_lineG_hypergraph}We store a hypergraph by its unsigned
incidence matrix, which is defined as an $N\times L$ matrix $R$ with the
entries $r_{j_{1}i}=r_{j_{2}i}=\cdots=r_{j_{k}i}=1$ and the other entries of
the $i$th column being $0$, when the hypergraph $i$ is incident to nodes
$j_{1},j_{2},\cdots,j_{k}$.

\begin{definition}
The line graph of a linear hypergraph $H\left(  N,L\right)  $ is defined as a
graph $l\left(  H\right)  $, of which the node set is the set of the
hyperlinks of the hypergraph and two nodes are connected by an unweighted link
when the corresponding hyperlinks share one node.
\end{definition}

\begin{definition}
The line graph of a nonlinear hypergraph $H\left(  N,L\right)  $ is defined as
a graph $l\left(  H\right)  $, of which the node set is the set of the
hyperlinks of the hypergraph and two nodes are connected by an link of weight
$t$ when the corresponding hyperlinks share $t$ node(s).
\end{definition}

We observe that the line graph $l\left(  H\right)  $ is a simple graph when
$H\left(  N,L\right)  $ is linear, and $l\left(  H\right)  $ of nonlinear
hypergraph $H\left(  N,L\right)  $ is a weighted graph. The adjacency matrix
of the line graphs of hypergraphs can be computed from the unsigned incidence
matrices of hypergraphs.

\begin{figure}
[ptb]
\begin{center}
\includegraphics[
height=2.8089in,
width=3.128in
]%
{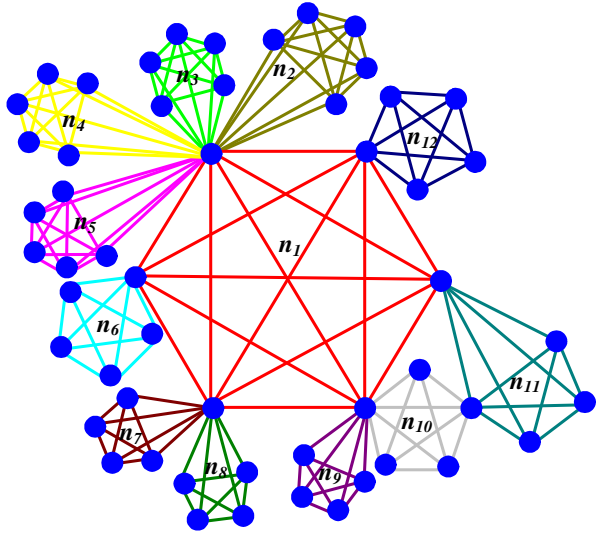}%
\caption{The line graph of the hypergraph in Figure
\ref{hypergraph_socialnetworks}. The nodes here denote the individuals while
the communities consist of links of the same color and the shared thick black
link(s) and the nodes incident to the links.}%
\label{linegraph_hypergraphsocialnetworks}%
\end{center}
\end{figure}

In Figure \ref{linegraph_hypergraphsocialnetworks} we show the line graph of
the hypergraph of Figure \ref{hypergraph_socialnetworks}. As depicted, there
are $12$ communities, of which $5$ communities have $6$ members and $7$
communities have $5$ members. Table \ref{Tbl_communities_nodes_linegraph}
shows the members of all the communities of the network in Figure
\ref{linegraph_hypergraphsocialnetworks}. We see that the line graph display
the community structure and the overlap better.%

%TCIMACRO{\FRAME{ftbpFU}{3.128in}{2.8089in}{0pt}{\Qcb{The line graph of the
%hypergraph in Figure \ref{hypergraph_socialnetworks}. The nodes here denote
%the individuals while the communities consist of links of the same color and
%the shared thick black link(s) and the nodes incident to the links.}%
%}{\Qlb{linegraph_hypergraphsocialnetworks}}%
%{linegraph_hypergraphSocialnetworks_.eps}%
%width 3.128in;  height 2.8089in;  depth 0pt;  original-width 3.2059in;
%original-height 3.2508in;  cropleft "0";  croptop "1";  cropright "1";
%filename 'linegraph_hypergraphSocialnetworks_.eps';file-properties "XNPEU";}%
%}}%
%BeginExpansion
%EndExpansion

\section{The relation between the maximum overlapping depth $k_{\max}$ and the
smallest adjacency eigenvalue of the corresponding line graph}

\subsection{The line graph of linear and $k$-uniform hypergraph $H_{k}\left(
N,L\right)  $}

Since $H_{k}\left(  N,L\right)  $ is $k$-uniform, the unsigned incidence
matrix $R$ of $H_{k}\left(  N,L\right)  $ has exactly $k$ $1$-entries and
$N-k$ $0$-entries in each column, and we have $k_{\max}=k$. Hence, all the
diagonal entries of $R^{T}R$ are $k$. Due to the definition of linearity of
hypergraphs, two columns of $R$ of $H_{k}\left(  N,L\right)  $ have at most
one $1$-entry at the same row. Hence, all the non-diagonal entries of $R^{T}R$
are either $1$ or $0$. In addition, $R^{T}R$ is a Gram matrix
\cite{PVM_SpectraCUP,Cvetkovic_eig_unsignLaplacian}. Therefore the adjacency
matrix of the line graph of linear and $k$-uniform hypergraph $H_{k}\left(
N,L\right)  $ is,%
\begin{equation}
A_{l\left(  H_{k}\right)  }=R^{T}R-kI \label{Eq_adj_matrix_line_H}%
\end{equation}

Both of the matrices $\left(  R^{T}R\right)  _{L\times L}$ and $\left(
RR^{T}\right)  _{N\times N}$ are positive semidefinite,%
\begin{align*}
x^{T}\left(  R^{T}R\right)  x  &  =\left(  Rx\right)  ^{T}Rx=\left\Vert
Rx\right\Vert _{2}^{2}\geq0\\
x^{T}\left(  RR^{T}\right)  x  &  =\left(  R^{T}x\right)  ^{T}R^{T}%
x=\left\Vert R^{T}x\right\Vert _{2}^{2}\geq0
\end{align*}
All eigenvalues of $\left(  R^{T}R\right)  _{L\times L}$ are non-negative. Due
to $\left(  \ref{Eq_adj_matrix_line_H}\right)  $, the adjacency eigenvalues of
the line graph of linear and $k$-uniform hypergraph $H_{k}\left(  N,L\right)
$ are not smaller than $-k$, where $k$ is the overlapping depth.

We have more results for linear and uniform networks.

\begin{lemma}
[see \cite{PVM_SpectraCUP}]\label{lemma_eigval}For all matrices $A_{n\times
m}$ and $B_{m\times n}$ with $n\geq m$, it holds that $\lambda\left(
AB\right)  =\lambda\left(  BA\right)  $ and $\lambda\left(  AB\right)  $ has
$n-m$ extra zero eigenvalues,%
\[
\lambda^{n-m}\det\left(  BA-\lambda I\right)  =\det\left(  AB-\lambda
I\right)
\]

\end{lemma}

Using Lemma \ref{lemma_eigval} we have,%
\[
\det\left(  \left(  R^{T}R\right)  _{L\times L}-\lambda I\right)
=\lambda^{L-N}\det\left(  \left(  RR^{T}\right)  _{N\times N}-\lambda
I\right)
\]
Using the definition of the adjacency matrix of the line graph in $\left(
\ref{Eq_adj_matrix_line_H}\right)  $\ yields,%
\[
\det\left(  A_{l\left(  H_{k}\right)  }-\left(  \lambda-k\right)  I\right)
=\lambda^{L-N}\det\left(  \left(  RR^{T}\right)  _{N\times N}-\lambda
I\right)
\]
or%
\begin{equation}
\det\left(  A_{l\left(  H_{k}\right)  }-\lambda I\right)  =\left(
\lambda+k\right)  ^{L-N}\det\left(  \left(  RR^{T}\right)  _{N\times
N}-\left(  \lambda+k\right)  I\right)
\end{equation}
The adjacency matrix $A_{l\left(  H_{k}\right)  }$ has at least $L-N$
eigenvalues of $-k$, where $N$ is the number of communities and $L$ is the
number of individuals. The matrix $RR^{T}$ is positive semidefinite, hence,
the remaining $N$ eigenvalues of $A_{l\left(  H_{k}\right)  }$ are not smaller
than $-k$.

\subsection{The line graph of linear and non-uniform hypergraph $H\left(
N,L\right)  $ with $k_{\max}$}

Since the maximum overlapping depth of $H\left(  N,L\right)  $ is $k_{\max}$,
the unsigned incidence matrix $R$ of $H_{k}\left(  N,L\right)  $ has at most
$k_{\max}$ $1$-entries in each column. Therefore, the largest diagonal entry
of $R^{T}R$ is $k_{\max}$. The adjacency matrix of the line graph of a linear
and non-uniform hypergraph $H\left(  N,L\right)  $ is,%
\begin{equation}
A_{l\left(  H\right)  }=R^{T}R+C-k_{\max}I
\label{Eq_adj_matrix_lineG_linear_nonuniform}%
\end{equation}
where $C=\operatorname*{diag}\left(
\begin{array}
[c]{cccc}%
c_{11} & c_{22} & \cdots & c_{LL}%
\end{array}
\right)  $ and $c_{jj}\geq0$, $1\leq j\leq L$. By adding $C$ to $R^{T}R$, we
make all the diagonal entries of $R^{T}R+C$ equal to $k_{\max}$.

We show that $R^{T}R+C$ is also positive semidefinite.%
\begin{align}
x^{T}\left(  R^{T}R+C\right)  x  &  =x^{T}\left(  R^{T}R\right)
x+x^{T}\left(  \sqrt{C}^{T}\sqrt{C}\right)  x\nonumber\\
&  =\left\Vert Rx\right\Vert _{2}^{2}+\left\Vert \sqrt{C}x\right\Vert _{2}%
^{2}\geq0
\end{align}
where $x_{L\times1}$ is an arbitrary vector and $\sqrt{C}=\operatorname*{diag}%
\left(
\begin{array}
[c]{cccc}%
\sqrt{c_{11}} & \sqrt{c_{22}} & \cdots & \sqrt{c_{LL}}%
\end{array}
\right)  $. Hence, the adjacency eigenvalues of the line graph of a linear and
non-uniform hypergraph $H\left(  N,L\right)  $ are not smaller than $-k_{\max
}$, where $k_{\max}$ is the maximum overlapping depth of $H\left(  N,L\right)
$.

\subsection{The line graph of nonlinear and non-uniform hypergraph $H\left(
N,L\right)  $ with $k_{\max}$}

Since $H\left(  N,L\right)  $ is nonlinear, there are some pairs of hyperlinks
sharing more than one nodes. If hyperlink $i$ and hyperlink $j$ share $t$
nodes, then, by the definition of the line graph of hypergraph $H\left(
N,L\right)  $, the link weight of the corresponding link between node $i$ and
$j$ in the line graph is $t$. The line graph of nonlinear hypergraph $H\left(
N,L\right)  $ becomes a weighted graph. In the language of social networks,
the link weight of two individuals is $t$ if the two individuals are both
members of $t$ communities. The adjacency matrix of the line graph of
nonlinear and non-uniform hypergraph $H\left(  N,L\right)  $ is,
\[
A_{l\left(  H\right)  }=R^{T}R+C-k_{\max}I
\]
where $C=\operatorname*{diag}\left(
\begin{array}
[c]{cccc}%
c_{11} & c_{22} & \cdots & c_{LL}%
\end{array}
\right)  $ and $c_{jj}\geq0$, $1\leq j\leq L$. By adding $C$ to $R^{T}R$, we
make all the diagonal entries of $R^{T}R+C$ are $k_{\max}$. We have proved
that $R^{T}R+C$ is positive semidefinite, hence, the adjacency eigenvalues of
the line graph of nonlinear and non-uniform hypergraph $H\left(  N,L\right)  $
are also not smaller than $-k_{\max}$.

\section{Hypergraphs with power-law degree distribution}

As a common property, the node degree of many real-world large networks
including social networks follows a power-law distribution
\cite{Albert:statistical_mechanics_networks,BApowerlaw}. To model social
networks better, we need to incorporate the power-law degree distribution into
our hypergraph model. We introduce network growing and preferential attachment
to our hypergraph model.

\begin{figure}
[ptb]
\begin{center}
\includegraphics[
height=1.4373in,
width=2.9317in
]%
{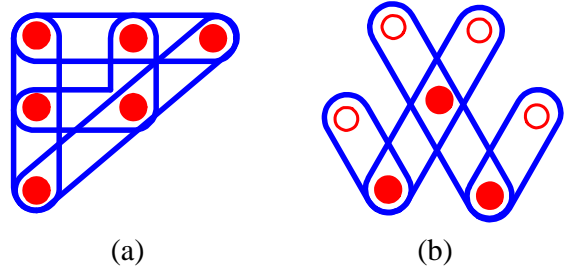}%
\caption{ The seed (the starting hypergraph) (a) and the growing element (b)
we use in the simulation. At each time step a growing element is added to the
existing hypergraph. All the hyperlinks of the growing elements has only one
red circle means that they can only connect to one more node.}%
\label{hypergraph_preferential_attachment}%
\end{center}
\end{figure}

By preferential attachment, we generate linear and non-uniform hypergraphs
only with overlapping depth of $2$ and $3$. Starting with a small hypergraph
(with $m_{0}$ nodes, $m_{0}>4$), which we call as a seed, at every time step
we add a growing element which consists of three nodes and two hyperlinks of
overlapping depth of $2$ and two hyperlinks of overlapping depth of $3$. The
four hyperlinks connect all the three nodes of a growing element to the
existing hypergraph. Note that all four hyperlinks can only connect to one
more node. The probability $\Pi$ that a hyperlink will connect to a node $i$
depends on the current degree $S_{i}$ of $i$, $\Pi\left(  i\right)
=S_{i}/\sum S_{i}$, where $\sum S_{i}$ is the sum of degrees of all the
existing nodes. In order to guarantee the linearity, the four hyperlinks must
connect to different existing nodes at each time step. Figure
\ref{hypergraph_preferential_attachment} shows us the seed and the growing
element we use in the simulation.
%TCIMACRO{\FRAME{ftbpFU}{2.9317in}{1.4373in}{0pt}{\Qcb{ The seed (the starting
%hypergraph) (a) and the growing element (b) we use in the simulation. At each
%time step a growing element is added to the existing hypergraph. All the
%hyperlinks of the growing elements has only one red circle means that they can
%only connect to one more node.}}{\Qlb{hypergraph_preferential_attachment}%
%}{hypergraph_preferential_attachment.eps}%
%width 2.9317in;  height 1.4373in;  depth 0pt;  original-width 3.416in;
%original-height 1.6587in;  cropleft "0";  croptop "1";  cropright "1";
%filename 'hypergraph_preferential_attachment.eps';file-properties "XNPEU";}}}%
%BeginExpansion
%EndExpansion%
%TCIMACRO{\TeXButton{B}{\begin{table}[tbp] \centering}}%
%BeginExpansion
\begin{table}[tbp] \centering
%EndExpansion%
\begin{tabular}
[c]{|l|l|l|l|l|l|l|l|}\hline
Network & $N$ & $L$ & $\alpha$ & $C$ & $C_{r}$ & $\rho_{D}$ & $l$\\\hline
The Hyves social network & $10326$ & $1260458$ & $-0.88$ &
\multicolumn{1}{|l|}{$0.84$} & \multicolumn{1}{|l|}{$0.024$} & $0.29$ &
$6.7$\\\hline
The line graph of the & $1510$ & $32031$ & $-0.76$ & $0.58$ & $0.029$ & $0.72$
& $4.8$\\
generated hypergraph &  &  &  &  &  &  & \\\hline
\end{tabular}
%TCIMACRO{\TeXButton{caption}{\caption
%{The properties of an social network retrieved from Hyves and the line graph of the hypergraph generated by our hypergraph model. The properties measured are: the total number of nodes $N$, the total number of nodes $L$, exponent $\alpha
%$ of the power-law degree distribution, clustering coefficient $C$, assortativity coefficient $\rho
%_{D}$ (we have employed the formula in \cite{Piet:influence_assortativity}%
%), average path length $l$. For a comparison we have included the clustering coefficient $C_{r}%
%$ of a ER random graph with the same size and link density.}}}%
%BeginExpansion
\caption
{The properties of an social network retrieved from Hyves and the line graph of the hypergraph generated by our hypergraph model. The properties measured are: the total number of nodes $N$, the total number of nodes $L$, exponent $\alpha
$ of the power-law degree distribution, clustering coefficient $C$, assortativity coefficient $\rho
_{D}$ (we have employed the formula in \cite{Piet:influence_assortativity}%
), average path length $l$. For a comparison we have included the clustering coefficient $C_{r}%
$ of a ER random graph with the same size and link density.}%
%EndExpansion
\label{Tbl_properties_comparison}%
%TCIMACRO{\TeXButton{E}{\end{table}}}%
%BeginExpansion
\end{table}%
%EndExpansion
$\rho_{D}$

\begin{figure}
[th]
\begin{center}
\includegraphics[
height=3.9115in,
width=4.4564in
]%
{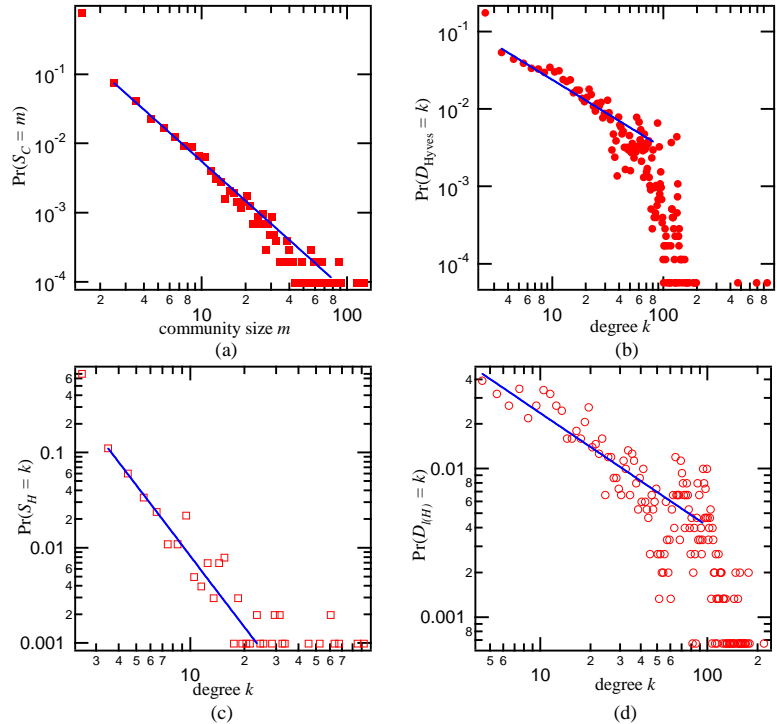}%
\caption{The community size distribution of the Hyves social network (a) and
the degree distribution of the generated hypergraph (c), and the degree
distribution of the Hyves social network (b) and the degree distribution of
the line graph of the generated hypergraph (d). They are all fitted by
power-law function $f(x)=\beta x^{\alpha}$, and $\alpha=-1.88$ (a),
$\alpha=-2.5$ (c), $\alpha=-0.88$ (b), $\alpha=-0.76$ (d).}%
\label{Fig_power_law_comparison}%
\end{center}
\end{figure}

Using this model (with the seed and the growing element in Figure
\ref{hypergraph_preferential_attachment}), we generate a hypergraph $H$ with
$1015$ nodes and $1510$ hyperlinks, which is stored in the unsigned incidence
matrix $R$. By the formula $\left(
\ref{Eq_adj_matrix_lineG_linear_nonuniform}\right)  $, we compute the
adjacency matrix of the line graph $l\left(  H\right)  $. The line
graph\footnote{This line graph is unweighted, since the hypergraph we have
generated is linear.} of the generated hypergraph has $1510$ nodes and $32031$
links. The degree $D_{H}$ of a random node of a hypergraph is defined as the
number of hyperlinks which are incident to that node, and it is essentially
equal to the size of the corresponding community. The degree distribution
$\Pr\left(  D_{H}=k\right)  $ of that generated hypergraph denotes actually
the community size distribution, and strictly follows power-law distribution.
The degree of a random node of the line graph is denoted as $D_{l\left(
H\right)  }$, and we show in Figure \ref{Fig_power_law_comparison} that the
degree distribution $\Pr\left(  D_{l\left(  H\right)  }=k\right)  $ of the
line graph approximately follows power-law distribution.

As the most popular online social networking site in Netherlands, Hyves has
more than 10 million users, which means that more than half of the Dutch
population are using Hyves. Nearly half of Hyves users make their profiles
open to the public. From the open profiles we can see some information of
users including companies, schools, colleges, clubs and other organizations,
to which they belong. By using a breath-first search we found out that there
are $17619$ users claiming that they belong to some communities. The total
number of these communities are $10326$. We make a network with $17619$ users
as nodes, and two users are connected by a link when they belong to the same
community. We denote the size of a community as $S_{C}$, which is defined as
the total number of individuals belonging to that community. We compute the
properties of the Hyves social network and the line graph of the hypergraph
generated by our hypergraph model with preferential attachment. As shown in
Table \ref{Tbl_properties_comparison}, both of these two networks have high
clustering coefficient, positive assortativity coefficient, short average path
length and similar exponent of the power-law degree distribution, although the
size of the line graph is much smaller than the size of the Hyves social
network. As depicted in Figure \ref{Fig_power_law_comparison} (a) and (b), the
community size of the Hyves social network follows a power-law distribution
with exponent $\alpha=-1.88$, and the degree distribution of the Hyves social
network can also be fitted by a power-law function with exponent
$\alpha=-0.88$. Figure \ref{Fig_power_law_comparison} (c) and (d) show us that
the power-law degree distribution of the generated hypergraph $\alpha=-2.5$ is
quite similar with that of the community size distribution of Figure
\ref{Fig_power_law_comparison} (a), and the exponent of power-law degree
distribution of the line graph $\alpha=-0.76$ seems very close to the exponent
in Figure \ref{Fig_power_law_comparison} (b). Table
\ref{Tbl_properties_comparison} and Figure \ref{Fig_power_law_comparison} show
that our hypergraph model with preferential attachment has the common
properties of real-world social networks, besides that community structure and
community overlap are already incorporated.
%TCIMACRO{\FRAME{fthFU}{4.4564in}{3.9115in}{0pt}{\Qcb{The community size
%distribution of the Hyves social network (a) and the degree distribution of
%the generated hypergraph (c), and the degree distribution of the Hyves social
%network (b) and the degree distribution of the line graph of the generated
%hypergraph (d). They are all fitted by power-law function $f(x)=\beta
%x^{\alpha}$, and $\alpha=-1.88$ (a), $\alpha=-2.5$ (c), $\alpha=-0.88$ (b),
%$\alpha=-0.76$ (d).}}{\Qlb{Fig_power_law_comparison}}%
%{power_law_comparison.eps}{\special{ language "Scientific Word";
%type "GRAPHIC";  maintain-aspect-ratio TRUE;  display "USEDEF";
%valid_file "F";  width 4.4564in;  height 3.9115in;  depth 0pt;
%original-width 5.9049in;  original-height 5.1776in;  cropleft "0";
%filename 'power_law_comparison.eps';file-properties "XNPEU";}}}%
%BeginExpansion
%EndExpansion

\section{Conclusion}

We have modeled social networks with overlapping communities by hypergraphs
and the line graphs of hypergraphs. The hyperlinks and nodes represent the
individuals and the communities respectively. If an individual belongs to
several communities, the corresponding nodes are connected by the
corresponding hyperlink. Since the line graphs of hypergraphs are just simple
graphs or weighted graph, we can implement the current network analysis
techniques. We defined the overlapping depth $k$ of an individual by the
number of communities that overlap in that individual, and we proved that the
minimum adjacency eigenvalue of the line graphs of hypergraphs is not smaller
than $-k_{\max}$, which is the maximum overlapping depth of the whole network.
We established a network model which incorporates overlapping communities
structures for the first time with tunable overlapping parameters. By
comparing our model with the online social network Hyves, we have shown that
our network model possesses the common properties of large social networks.

\bibliographystyle{splncs}
\bibliography{Dajie_linegraph}

\appendix{}

\end{document}